\documentclass[conference]{IEEEtran}
\IEEEoverridecommandlockouts
\usepackage{cite}
\usepackage{amsmath,amssymb,amsfonts}
\usepackage{algorithmic}
\usepackage{algorithm}
\usepackage{graphicx}
\usepackage{textcomp}
\usepackage{xcolor}
\usepackage{balance}
\def\BibTeX{{\rm B\kern-.05em{\sc i\kern-.025em b}\kern-.08em
    T\kern-.1667em\lower.7ex\hbox{E}\kern-.125emX}}
    
\begin{document}

\title{IVACS: \underline{I}ntelligent \underline{V}oice \underline{A}ssistant for \underline{C}oronavirus Disease (COVID-19) \underline{S}elf-Assessment\\
}

\makeatletter
\newcommand{\linebreakand}{%
  \end{@IEEEauthorhalign}
  \hfill\mbox{}\par
  \mbox{}\hfill\begin{@IEEEauthorhalign}
}
\makeatother

\author{\IEEEauthorblockN{Parashar Dhakal}
\IEEEauthorblockA{\textit{Electrical Controls Engineer} \\
\textit{Grote Industries}\\
Madison, IN, USA \\
Parashar.Dhakal@grote.com}
\and
\IEEEauthorblockN{Praveen Damacharla}
\IEEEauthorblockA{\textit{CTO and Research Scientist} \\
\textit{KineticAI Inc.,}\\
Crown Point, IN, USA \\
Praveen@KineticAI.com}\\
\and
\IEEEauthorblockN{Ahmad Y. Javaid}
\IEEEauthorblockA{\textit{EECS Department} \\
\textit{The University of Toledo}\\
Toledo, OH, USA \\
Ahmad.Javaid@Utoledo.edu}

\linebreakand
\centering
\IEEEauthorblockN{Hari K. Vege}
\IEEEauthorblockA{\textit{CSE Department} \\
\textit{Koneru Lakshmaiah University}\\
Vaddeswaram, AP, India \\
Hari.Vege@kluniversity.in}
\and
\IEEEauthorblockN{Vijay K. Devabhaktuni} 
\IEEEauthorblockA{\textit{ECE Department} \\ 
\textit{Purdue University Northwest}\\
Hammond, IN, USA\\
Vjdev@pnw.edu}
}

\maketitle

\begin{abstract}
At the time of writing this paper, the world has around eleven million cases of COVID-19, scientifically known as severe acute respiratory syndrome corona-virus 2 (SARS-COV-2). One of the popular critical steps various health organizations are advocating to prevent the spread of this contagious disease is self-assessment of symptoms. Multiple organizations have already pioneered mobile and web-based applications for self-assessment of COVID-19 to reduce the spread of this global pandemic. We propose an intelligent voice-based assistant for COVID-19 self-assessment (IVACS). This interactive assistant has been built to diagnose the symptoms related to COVID-19 using the guidelines provided by the Centers for Disease Control and Prevention (CDC) and the World Health Organization (WHO). The empirical testing of the application has been performed with 22 human subjects, all volunteers, using the NASA Task Load Index (TLX), and subjects’ performance accuracy has been measured. The results indicate that the IVACS is beneficial to users. However, it still needs additional research and development to promote its widespread application.

\end{abstract}


\section{Introduction}
The novel coronavirus (COVID-19) was first observed in late 2019 in Wuhan, China, and the patients suffered from a form of pneumonia \cite{b1}. The virus was identified as genus beta-coronavirus, placing it in the same category as the previously discovered deadly viruses such as Severe Acute Respiratory Syndrome (SARS) and Middle East Respiratory Syndrome (MERS). The virus has now spread across more than 200 countries. The WHO declared this virus a global health emergency in early 2020. National emergency was declared in the US in March 2020. More than 540,000 people have died from this virus across the globe, with more than 130,000 deaths in the US alone as of July 5, 2020 \cite{b2}. 

Hospitals and clinics around the world have been over- whelmed with the cases of COVID-19. With a lot of panic and rumors, people are visiting clinics and hospitals for other non-related symptoms. These visits are causing increased healthcare costs and spread of infection, while overloading the healthcare system. Self-assessment is therefore being studied as one of the solutions to this problem. This technique, i.e. self-assessment, has been used in the field of healthcare for a long time as it helps in learning, functioning more effectively, and fostering self-agency and authority \cite{b3,b4}.

Some recent examples of applications that have been built to self- assess the COVID-19 are \cite{b5,b6}. Though these applications are beneficial, they might not be accessible to all. Moreover, such apps are not useful for someone who does not know how to read, use a computer, or is visually impaired. Therefore, in this paper, we propose a novel idea to use the IVACS for self-assessment of the COVID-19. This interactive application, based on medical condition, helps in more precise clinical decision making about seeking medical care or taking rest at home; without burdening the hospitals at these challenging times. It also educates people with critical information. Primary contributions of this proposed work are below:
\begin{itemize}
 \item A new IVACS architecture for the self-assessment of COVID-19,
 \item A detailed study on the performance of the proposed IVACS,
 \item A study on the performance of the user in cohesion with IVACS, and
 \item A measurement of the perceived mental overload in user due to IVACS. 
\end{itemize}

\begin{figure*}[h]
 \centering
 \includegraphics[width=1\linewidth]{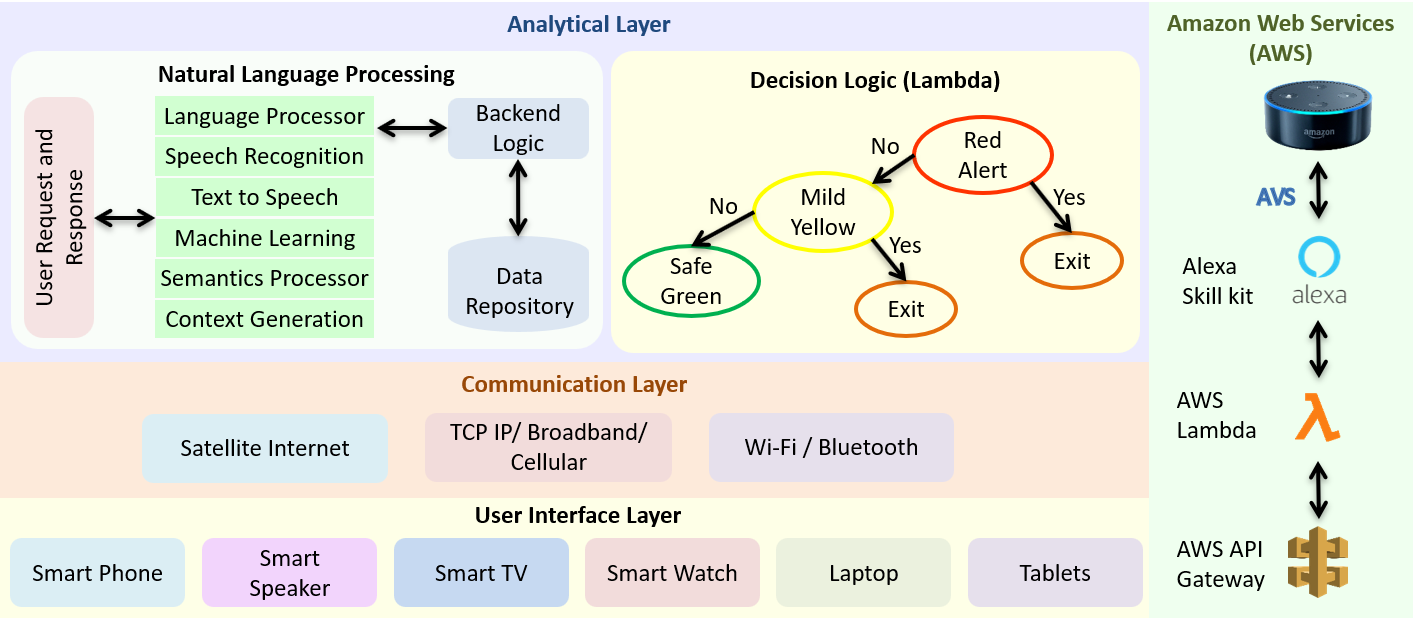}
  \caption{Proposed Generalized Architecture of the IVACS along with the Cloud Based Model}
  \label{arch}
\end{figure*}

\section{Related Work}
Long before COVID-19 pandemic, a wearable healthcare assistant was developed to record contextual and physiological information \cite{b7}. This prototype named LifeMinder was used in sensing pulse waves, user’s actions/postures, capturing contextual photos, and continuous voices. Collected data was sent to a healthcare PC and was retrieved on a web page for user accessibility. Some researchers, introduced HealthPal, an intelligent dialogue-based personal medical assistant, for self-monitoring of health \cite{b8}. This software was designed to help older citizens in monitoring their health without assistance. Researchers also developed an interactive robotic assistant for interacting with patients, measuring vital signs, and recording data \cite{b9}. The robot was interfaced with a blood pressure monitor and had a 3D face capable of displaying different emotions. Their initial study on interaction of patients with the proposed robotic assistant showed the performance improvement of the assistant because of interactivity involved in the task completion when they worked as a team. 

During early 2000s, most healthcare assistants were focused on the use of wearable devices and computer-based software applications. With the recent advancement of the artificial assistant, researchers have started to explore use of voice technology capable of decision making as a healthcare assistant. Some researchers developed patient-focused voice and web services using amazon Alexa and google assistant to solve the problem that the patients face using wearable health sensors \cite{b10}. The developed assistant was also capable of making suggestions, scheduling doctor appointments, and reminding the patient before therapy and appointments. Other researchers worked on the development of a voice-based assistant using amazon Alexa to help medical first responders in the treatment process \cite{b11}. They analyzed the developed assistant performance for a selected emergency treatment scenario where their result showed that the performance of care providers increases with the use of such assistant. All these past assistants were “task-specific”. In line with the previous works, our IVACS that uses amazon Alexa as a voice-based assistant for the self-assessment of COVID-19 based on CDC and WHO guidelines.

\section{Proposed Architecture}
Fig. \ref{arch} shows the overview of proposed IVACS architecture. The architecture is composed of 3 essential layers, namely user interface, communication, and analytical. The user interface layer consists of different hardware devices and components to interact with users such as smartphones, smart speakers, laptops, tablets, smart TVs, and Echo, where the input can be in any form such as text input or spoken language. Communication layers consist of network and protocols such as smart internet, Wi-Fi, Bluetooth, broadband, and cellular that can be used in connecting the hardware devices from user interface layers to the cloud platform in analytical layers. And the analytical layer consists of two blocks, namely Natural Language Processing (NLP) and decision logic block. The combination of NLP to the decision logic block in the cloud is one of the novel features of our architecture.
\begin{figure*}[h]
 \centering
 \includegraphics[width=0.95\linewidth]{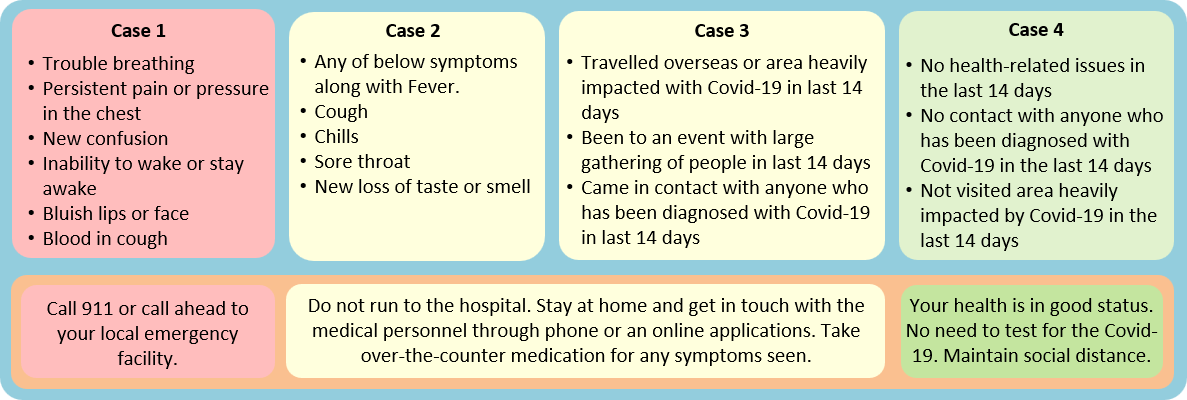}
 \caption{Guidelines followed for the self-assessment protocol}
 \label{guide}
\end{figure*} 

The process starts with the user requesting the hardware devices in the user interface layer through voice utterance. The request then passes on to the communication layer, and a voice-based input is received by a language processor in the analytical layer. Inside the language processor, speech recognition translates the utterance to text. This process is also called speech-to-text (STT) conversion that provides the text output to help the computer in processing the speech command. Similarly, semantics processor and context generator blocks help in further processing the text to interpret and understand user commands. Afterward, the command is passed to the backend block that contains the developed decision logic. Backend also communicates with the database repository to exchange stored information. After processing in the back-end, the response is sent to the language processor block again where the text-to-speech (TTS) conversion occurs, and the response in the form of speech is sent back to the user.

Our proposed IVACS was built into Amazon Web Services (AWS) and the primary block for our decision logic was developed inside the AWS lambda function using node.js. The user can access the proposed IVACS using amazon echo or the Amazon Alexa application. The information flow for our design in AWS has been depicted on the right side of Fig. \ref{arch}. The process starts by calling the wake word "ask Coronavirus," which signals the amazon echo to record the command to Alexa Voice Services (AVS). Next, Alexa Skills Kit (ASK) translates the voice command to text, and if the text matches a predefined utterance then, it calls the corresponding mapped function for that utterance in the lambda function using an intent request. After the execution of the intent, the response is returned to the ASK and then to the user. The lambda also communicates with the API gateway that lets us create and access an API. Besides, API gateway helps one to interact with the databases and messaging services through a secure gateway \cite{b16,b12}.

\section{Methodology}
Our experiment's main objective was to study the performance of the proposed IVACS and the participants individually and as a team for the self-assessment of COVID-19. During the experiment, the IVACS will interact with participants and guide them through the process upon knowing their health status. To make the interaction more effective and user-friendly, IVACS has been programmed in a way that it even gives an option of what to answer for each question it asks. In the process of experiment, different parameters such as errors committed by participants and IVACS, the number of interaction between IVACS and participants, the effect of IVACS on participant's performance, and total testing time were measured. In this section, we also presented the CDC and WHO recommended protocol to follow in the self-assessment of COVID-19 based on which the IVACS decision logic was built. Additionally, we discussed the participants' poll and different data collection methods employed to collect various parameters required for performance evaluation.

\subsection{COVID-19 Self Assessment Protocol}
In this section, we discuss the different cases considered in our application and subsequent recommendations made to the users based on their input to the IVACS. As per the CDC and WHO recommendations, the structure of guidelines that we followed in our application has been depicted in Fig. \ref{guide} \cite{b13,b14}. To provide better recommendations to the user, we divided the different conditions identified till date into three categories: red alert, mild yellow, and safe green. Users facing any of the symptoms falling under the red alert category were recommended to call 911 and visit the emergency immediately. Similarly, any users facing symptoms listed in the mild yellow were urged not to rush to the hospital and stay home, get in touch with the medical personnel through phone or online applications, and take over-the-counter medication. Besides, if the users had recently visited an area heavily impacted with COVID-19, a large gathering of people, or had been in contact with anyone diagnosed with COVID-19, they were recommended to stay in quarantine and get in touch with medical personnel through phone or online applications for possible suggestions. Finally, if the user does not fall under any of the two categories mentioned above, then they were declared safe and recommended to maintain social distance.

\subsection{Participants}
The participants for the experiment were the general population with no medical background at all. A total of 22 participants participated in the experiment belonging from different countries such as Nepal, the USA, India, and Bangladesh. The participants' age involved in this investigation ranged from 20 to 65 years and had an educational background ranging from high school to Ph.D. Among 22 participants, 15 were male and seven female.

\subsection{Experimental Setup}
For the experiment, we followed the guideline from our previous research publication, where we used the virtual assistant to help medical first responders in the treatment process \cite{b11}. Amazon Echo, or an Alexa application, was used in our experiment for interaction with participants after a comparison of various virtual assistants \cite{b15}. The experiment was carried out by sending the application to each participant. Participants were instructed to perform the test in a quiet environment in their homes, where they were monitored over a video call. As part of the experiment, a general 2-minute briefing of the experiment, testing, and survey was done before starting the experiment. However, none of the participants had any idea about the experiment before actually performing it and were only provided with the word "ask Coronavirus" to trigger the IVACS. Besides, any chance of the interaction between two fellow participants was forbidden during all stages to avoid human factor bias.

The experiment consisted of 18 execution steps with various conditions; however, the total number of execution steps followed by each participant varied on a case by case basis. Similarly, testing time could last anywhere between 25 seconds to 140 seconds based on user response to questions asked by IVACS. As depicted in Fig. \ref{guide}, the process begins with asking about red alert cases and would immediately stop and make a corresponding recommendation if any of the cases are seen. If a participant experiences none of the cases from the red alert zone, then it will move to the mild yellow zone and check for various conditions. Finally, if nothing is seen, then the IVACS will enter the safe green zone declaring the patient as safe. During the testing phase, participants got no assistance and were expected to perform alone with IVACS. After the testing process, participants were asked to complete a survey, and the TLX form was used for the procedure. 

\subsection{Data Collection}
The experimental data such as error committed by participants and IVACS during the testing process, the interaction between IVACS and participants, the effect of IVACS on participant's performance, and average testing time were measured and collected visually through a video call. Similarly, parameters such as participant's frustration level, effort, Mental Demand (MD), Physical Demand (PD), Temporal Demand (TD) as part of the TLX survey, and IVACS's performance such as response time, errors in operation,  were collected. These were received using messages and emails from the participants for analysis of cognitive overload of IVACS during task performance. A separate timer was implemented to report the average time taken by participants during the testing process.

\section{Result and Discussion}
As mentioned above, the main purpose of this investigation was to assess the performance of the proposed IVACS, along with measuring the effects of human factors on the IVACS. This was achieved through different parameters recorded during the experimental testing process, as mentioned in the data collection sub-section above. To assess the participant's workload during the experiment, we used six parameters of the TLX scale, namely MD, PD, TD, performance, effort, and frustration. Each of these parameters had an exclusive self-rating index ranging from 1 to 21 points at a 1 point per division rate. To assign self-rating points, the post-testing questionnaire was adopted as a user validation process. The overall workload was determined using equation ~\eqref{GrindEQ__1_} that provided a normalization. 
\begin{equation} 
\label{GrindEQ__1_} 
\begin{array}{c} 
NASA\ TLX\ Score\ (Overall\ workload\ score)\\ =  \ 21 - (MD\ +PD +TD +Performance + Effort\\ +\ Frustration\ )
\end{array}
\end{equation} 

\begin{figure}[t]
    \centerline{\includegraphics[width=0.95\linewidth]{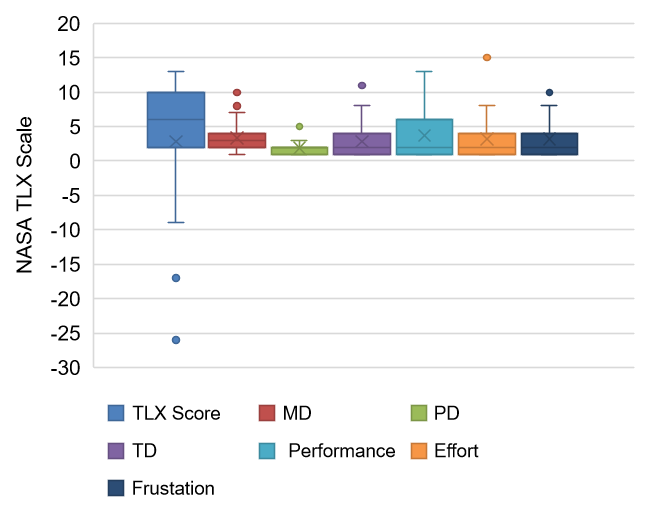}}
    \caption{The NASA task load index (NASA-TLX) ranking presentation}
    \label{fig1}
\end{figure}

Fig. ~\ref{fig1} shows the TLX ranking presentation where each parameter varied on a scale of 1 to 21 during the user survey. It shows the overall performance of participants using the TLX score, which has a range of -25 to 15 where 15 is the best performance, and any negative value was considered as a failed performance with the number of errors encountered during testing. We observed that many participants didn't feel any mental workload, or were overwhelmed by TD. However, some participants clearly showed frustration as they had to repeat some of the voice commands because of their diverse background and the accents of English they possessed. Our IVACS's NLP system also had a limitation in understating various dialects which we believe is one of the main reasons for some participants facing errors/ repetition in steps during self-diagnosis.
\begin{table}[!h]
\centering
\caption{\label{tab:table1} Average Time, Error(s), and Steps Executed}
\begin{tabular}{|c|c|c|c|}
    \hline
    \textbf{Parameters} & \textbf{Time(s)} & \textbf{Error(s)} & \textbf{Steps Executed}\\
    \hline
    Mean & 130 & 1 & 17 \\
    \hline
    Median & 120 & 0 & 18   \\
    \hline
     Mode  & 112 & 0 & 18 \\
    \hline
\end{tabular}
\end{table}
To evaluate the performance of both IVACS and participants as a team, we also computed the mean, median, and mode of time taken, error rate, and execution step for all the 22 experiments. The mean values for time, error, and action executed were found to be 130, 1, and 17, respectively, as shown in Table ~\ref{tab:table1}. To elaborate more, a mean execution step value of 17 and mode and median as 18, implies most of the participants for our experiment belonged to the safe green zone. In contrast, some belonged to the mild yellow zone. Similarly, a mean of 130 seconds and mode and median of 120 seconds indicates that some of the participants struggled in communication with IVACS and had to repeat speech utterance a couple of times to move to the next step. Besides, the mean error value was found to be 1. And the median and mode error values were zero, which signifies that there was some anomaly during the experiment, where for most of the participants, the IVACS worked well. However, some faced errors during the experiment, which increased the mean value. This error was noticed mostly seen among the non-native English speakers. In essence, IVACS showed different results for some of the
participants while it was uniform for most of them.
\begin{figure}[t]
    \centerline{\includegraphics[width=0.95\linewidth]{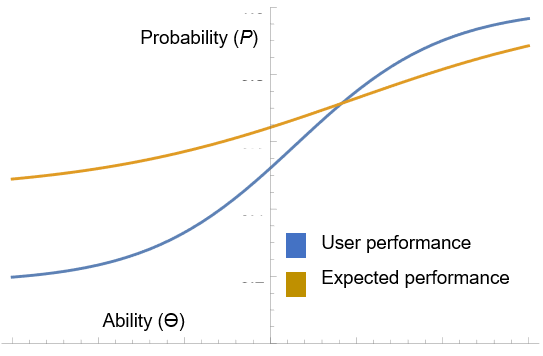}}
    \caption{Item Characteristic Curve for User Performance}
    \label{fig2}
\end{figure}

Besides, Item Characteristic Curve (ICC) was also used on a probabilistic scale to understand the relation between the expected performance versus the participant's actual performance for the given task ($\theta$). Fig. ~\ref{fig2} depicts the performance of participants during the experiment as compared to their expected performance based on the ability to perform on a probabilistic scale. It was observed for some participants' that the performance was slightly lower than the expected performance based on their ability ($\theta$). Various parameters might have caused this since, no formal training was required for this experimental testing and all the participants who took part in the experiment possessed the necessary skills to complete the given task. Our analysis of results and tests indicates the primary reason for some of the participants' lower performance during the experiment was due to different pronunciations and accents they possessed, which made it hard for the NLP system to recognize their speech utterance, resulting in an error/repetition during the experiment. 

\section{Conclusion and Future Work}
In this paper, we presented a novel real-time IVACS architecture for the self-assessment of COVID-19. The architecture was built inside AWS following the CDC and WHO guidelines. Besides, we also performed the empirical testing of the proposed architecture with 22 volunteers where we studied the performance accuracy of a proposed IVACS, the performance of the user in cohesion with IVACS, and the perceived mental overload in the user due to IVACS. The study of the perceived mental overload in users due to IVACS was done through the survey using the TLX form. As future work, we would like to include more volunteers for the experiment and study the response time of the proposed IVACS. Besides, we would also like to work on the performance improvement of IVACS and bring down the mean error value even for non-native speakers.


\balance

\bibliographystyle{IEEEtran}
\bibliography{Self_Assessment_CoronaVirus}

\end{document}